\documentclass[preprint,10pt,5p,twocolumn]{elsarticle}

\usepackage{preprintcover}  
\PreprintCoverPaperTitle{Measurement of angular correlations in Drell--Yan lepton pairs to probe \Zg\ boson transverse momentum at $\sqrt{s} =7$~TeV with the ATLAS detector}  
\PreprintIdNumber{CERN-PH-EP-2012-325}  
\PreprintCoverAbstract{A measurement of angular correlations in Drell--Yan lepton pairs via the \phist\ observable is presented. 
  This variable probes the same physics as the \Zg\ boson transverse momentum with a better experimental resolution. 
  The \Zgee\ and \Zgmm\ decays produced in proton--proton collisions at a centre-of-mass energy of $\rts = 7$~\TeV\ are used. 
  The data were collected with the ATLAS detector at the LHC and correspond to an integrated luminosity of $4.6$~\ifb. 
  Normalised differential cross sections as a function of $\phist$ are measured separately for electron and muon decay channels. 
  These channels are then combined for improved accuracy. The cross section is also measured double differentially as a function 
  of $\phist$ for three independent bins of the $Z$ boson rapidity. The results are compared to QCD calculations and to predictions 
  from different Monte Carlo event generators. The data are reasonably well described, in all measured $Z$ boson rapidity regions, 
  by resummed QCD predictions combined with fixed-order perturbative QCD calculations or by some Monte Carlo event generators. The measurement precision is typically better by one order of magnitude than present theoretical 
  uncertainties.}  
\PreprintJournalName{Phys.~Lett.~B}  

\usepackage{graphicx}
\usepackage{amssymb}

\usepackage{lscape}
\usepackage{subfigure}
\usepackage{rotating}
\usepackage{overpic}

\RequirePackage[colorlinks,breaklinks,pdftitle={ATLAS draft},pdfauthor={The ATLAS Collaboration}]{hyperref}
\hypersetup{linkcolor=blue,citecolor=blue,filecolor=black,urlcolor=blue}

\biboptions{sort&compress}

\usepackage{atlasphysics}
\usepackage{phistdefs}

\journal{Physics Letters B}
\renewcommand*{\today}{November 29, 2012} 

\begin{document}

\begin{frontmatter}

\title{Measurement of angular correlations in Drell--Yan lepton pairs to probe \Zg\ boson transverse momentum at $\sqrt{s} =7$~TeV with the ATLAS detector}
\author{The ATLAS Collaboration} 
\begin{abstract}
A measurement of angular correlations in Drell--Yan lepton pairs via the \phist\ observable is presented.
This variable probes the same physics as the \Zg\ boson transverse momentum with a better experimental resolution.
The \Zgee\ and \Zgmm\ decays produced in proton--proton collisions at a centre-of-mass energy of $\rts = 7$~\TeV\ are used. 
The data were collected with the ATLAS detector at the LHC and correspond to an integrated luminosity of $4.6$~\ifb.
Normalised differential cross sections as a function of $\phist$ are measured separately for electron and muon decay channels.
These channels are then combined for improved accuracy.
The cross section is also measured double differentially as a function of $\phist$ for three independent bins of the $Z$ boson rapidity.
The results are compared to QCD calculations and to predictions from different Monte Carlo event generators. 
The data are reasonably well described, in all measured $Z$ boson rapidity regions, by resummed QCD predictions combined with fixed-order 
perturbative QCD calculations or by some Monte Carlo event generators.
The measurement precision is typically better by one order of magnitude than present theoretical uncertainties.
\end{abstract}


\end{frontmatter}

\section{Introduction}

\par In hadron collisions at \TeV\ energies the vector bosons $W$ and $\Zg$ are 
copiously produced with non-zero momentum transverse to the beam direction ($p_{\rm T}$)
because of radiation of quarks and gluons from the initial-state partons. 
In this context the signatures \Zgee\ and \Zgmm\ provide an ideal testing ground 
for QCD due to the absence of colour flow between the initial and final state~\cite{Bozzi:2010xn,Mantry:2010bi,Becher:2010tm}. 
The study of the low \zpt\ spectrum (\zpt~\textless~\mZ), which dominates the cross section, 
has important implications on the understanding of Higgs boson production since the transverse-momentum resummation 
formalism required to describe the \Zg\ boson cross section is valid also for the Higgs boson~\cite{Bozzi:2003jy,Bozzi:2005wk,deFlorian:2011xf,Berge:2004nt}.
A precise understanding of the \zpt\ spectrum is also necessary to further improve 
the modelling of $W$ boson production in QCD calculations and Monte Carlo (MC) event generators, since the 
measurement of the $W$ mass is directly affected by uncertainties in the $p_{\rm T}^W$ shape~\cite{Aaltonen:2012bp,Abazov:2012bv}. 

\par The transverse momentum spectra of $W$ and $\Zg$ bosons produced via the Drell--Yan mechanism 
have been extensively studied by the Tevatron collaborations~\cite{Affolder:1999jh,Abbott:1999yd,Abazov:2007nt,Abazov:2010kn,Aaltonen:2012fi}
and, recently, also by the LHC experiments~\cite{Aad:2011gj,Chatrchyan:2011wt,Aad:2011fp}.
However, the precision of direct measurements of the \Zg\ spectrum at low \zpt\ at the LHC and the Tevatron is limited by the experimental 
resolution and systematic uncertainties rather than by the size of the available data samples.
This limitation affects
the choice of bin widths and the ultimate precision of the \zpt\ spectrum.
In recent years, additional observables with better experimental resolution 
and smaller sensitivity to experimental systematic uncertainties have been investigated~\cite{Maarten:2010,Vesterinen:2009,Banfi:2011,Banfi:2011dx}.
The optimal experimental observable to probe the low-\ptz\ domain of \Zg\ production was found to be $\phist$ which is defined~\cite{Banfi:2011} as:

\begin{equation}
\phist \equiv \tan(\phi_{\mathrm{acop}}/2) \cdot  \sin(\theta^{*}_{\eta}) \; ,
\label{eq:phist}
\end{equation}
\noindent 
where $\phi_{\mathrm{acop}} \equiv \pi - \Delta\phi$, $\Delta\phi$ being the azimuthal opening angle between 
the two leptons, and the angle $\theta^{*}_{\eta}$ is a measure of the  scattering angle of the leptons with respect to the proton beam direction in the rest frame of the dilepton system.
The angle $\theta^{*}_{\eta}$ is defined~\cite{Banfi:2011} by $\cos(\theta^{*}_{\eta}) \equiv \tanh[(\eta^{-} - \eta^{+})/2]$ where $\eta^{-}$ 
and $\eta^{+}$ are the pseudorapidities\footnote{ATLAS uses a right-handed
coordinate system with its origin at the nominal $pp$ interaction point (IP) in
the centre of the detector and the $z$-axis along the beam pipe. The $x$-axis points
from the IP to the centre of the LHC ring, and the $y$-axis points upward. Cylindrical
coordinates $(r,\phi)$ are used in the transverse plane, $\phi$ being the azimuthal
angle around the beam pipe. The pseudorapidity is defined in terms of the polar angle
$\theta$ as $\eta=-\ln\tan(\theta/2)$ and the rapidity is defined as $y=\ln[(E+p_{z})/(E-p_{z})] / 2$.}
 of the negatively and positively charged lepton, respectively. 
Therefore, $\phist$ depends exclusively on the directions of the two lepton tracks, which are better measured than their momenta.
The $\phist$ variable is positive by definition.
It is correlated to the quantity $p_{\rm T}^Z/m_{\ell\ell}$, where $m_{\ell\ell}$ is the invariant mass of the lepton pair, and therefore probes the same physics as the transverse momentum \zpt~\cite{Banfi:2012du}.
Values of $\phist$ ranging from $0$ to $1$ probe the $\zpt$ distribution mainly up to $\sim 100$~GeV.
The \phist\ distribution of $\Zg$ bosons has been measured in three bins of the $Z$ boson rapidity (\yZ) by the D\O~Collaboration using $7.3$~\ifb\ of $p\bar{p}$ collisions at $\sqrt{s} = 1.96 \TeV$~\cite{Abazov:2010mk}.
\par This Letter presents a measurement of the normalised \phist\ distribution in bins of the $Z$ boson rapidity \yZ\ using $4.6$~\ifb\ of $pp$ interactions 
collected at $\sqrt{s} =7$~TeV in $2011$ by the ATLAS detector. 
The normalised differential cross section is measured in both the electron 
and muon channels in the fiducial lepton acceptance defined by the lepton ($\ell$ = $e$, $\mu$) transverse momentum $p_{\rm T}^{\ell} > 20\GeV$, the 
lepton pseudorapidity $|\eta^{\ell}| < 2.4$ and the invariant mass of the lepton pair $66\GeV<m_{\ell\ell}<116\GeV$. 
Correction factors allowing the extrapolation of the cross section from the fiducial letpton acceptance to the full lepton acceptance, restricted to $66\GeV<m_{\ell\ell}<116\GeV$, are also presented.  
The reconstructed \phist\ distribution, after background subtraction, is corrected for all detector effects. 
The measurements are reported with respect to three distinct reference points at particle level regarding QED final-state radiation (FSR) corrections. 
The true dilepton mass $m_{\ell\ell}$ and $\phist$ are defined by the final-state leptons after QED FSR (``bare'' leptons), or by recombining them with 
radiated photons within a cone of $\Delta R = \sqrt{(\Delta \eta)^2 + (\Delta \phi)^2} = 0.1$ (``dressed'' leptons), or by the final-state leptons before 
QED FSR (``Born leptons''). 
The bare definition does not require any QED FSR correction for muons, whilst the dressed definition is the closest to the experimental measurement for electrons.
The Born definition corresponds to the full correction for QED FSR effects, so that it can be used for the combination of the 
electron and muon channels. 
The combination of the electron and muon channels is compared to QCD predictions obtained by matching resummed and fixed order QCD calculations, as well as to the predictions of MC event generators implementing a parton shower (PS) algorithm.
%
%
\section{QCD predictions}
\label{sec:Predictions}
%
%
\par Non-zero $\ptz$ is mainly generated through the emission of partons in the initial state.
In the high $\ptz$ region ($\ptz \gtrsim m_Z$) the spectrum is determined primarily by hard parton emission. 
Perturbative QCD calculations, based on the truncation of the 
perturbative series at a fixed order in $\alpha_s$, are theoretically justified and provide reliable predictions. 
The inclusive cross-section prediction is finite but the differential 
cross section diverges as $\zpt$ approaches zero. In this limit ($\zpt \ll m_Z$) the convergence
of the fixed-order expansion is spoiled by the presence of powers of large logarithmic terms 
which have to be resummed to restore the convergence. 
\par Differential cross
sections calculated to $\mathcal{O}(\alpha_s^2)$ are available for \Zg\ production through the 
{\sc Fewz}~\cite{Melnikov:2006kv,Gavin:2010az} and {\sc Dynnlo}~\cite{Catani:2009sm,Catani:2007vq} programs. 
The {\sc ResBos}~\cite{Ladinsky:1993zn,Balazs:1997xd,Landry:2002ix} generator resums the leading 
contributions up to next-to-next-to-leading logarithms (NNLL) and matches the result to fixed-order 
calculations at $\mathcal{O}(\alpha_s)$. This is corrected to $\mathcal{O}(\alpha_s^2)$ using a $k$-factor depending 
on \zpt\ and \yZ~\cite{Arnold198937}.
In addition, the {\sc ResBos} generator includes a non-perturbative form factor that needs to be determined from data~\cite{Landry:1999an}. 
A slightly different approach has been proposed recently to describe the Tevatron Run II data by matching NNLL accuracy 
to {\sc MCFM} calculations~\cite{Campbell:2002tg}, with no apparent need for non-perturbative contributions~\cite{Banfi:2011dm,Banfi:2012du}. 
\par Similarly to resummed calculations, PS algorithms 
such as those used in {\sc Pythia}~\cite{pythia} and {\sc Herwig}~\cite{herwig} provide an all-order approximation of parton radiation in the soft 
and collinear region through the iterative splitting and radiation of partons. 
The {\sc Powheg}~\cite{Nason:2004rx,powheg,Alioli:2010xd,Alioli:2008gx} and \mcatnlo~\cite{mcatnlo} event generators combine next-to-leading 
order (NLO) QCD matrix elements with a PS algorithm to produce differential cross-section predictions that are finite for all $\zpt$. 
The {\sc Alpgen}~\cite{Mangano:2002ea} and {\sc Sherpa}~\cite{Gleisberg:2008ta} event generators implement tree-level matrix elements 
for the generation of multiple hard partons in association with the weak boson. 
They are matched to parton showers either by a PS algorithm using re-weighting procedures~\cite{Hoeche:2009rj,Krauss:2002up} or through a 
veto~\cite{Mangano:2002ea}, in order to avoid the double counting of QCD emissions in the matrix element and the parton shower.

\section{The ATLAS detector}
\label{sec:Detector}

\par The ATLAS detector~\cite{DetectorPaper:2008} is a multi-purpose particle physics
detector operating at one of the beam interaction points of the LHC. It covers nearly the entire 
solid angle around the collision region and consists of an inner tracking detector (inner detector or ID) surrounded 
by a thin superconducting solenoid providing a $2$~T axial magnetic field, electromagnetic and hadronic 
calorimeters, and a muon spectrometer (MS).

\par Measurements in the ID are performed with silicon pixel 
and microstrip detectors covering $|\eta|< 2.5$. A straw-tube tracking detector follows radially and covers
the range $|\eta|< 2.0$. The lead/liquid-argon electromagnetic calorimeter is divided into 
barrel ($|\eta|< 1.5$) and endcap ($1.4< |\eta| < 3.2$) sections. The hadronic calorimeter is 
based on steel/scintillating tiles in the central region ($|\eta|< 1.7$), and is extended to $|\eta| = 4.9$ 
by endcap and forward calorimeters which use liquid argon. The MS comprises separate trigger 
and high-precision tracking chambers to measure the deflection of muons in a magnetic field generated by three 
large superconducting toroids arranged with an eightfold azimuthal coil symmetry around the calorimeters. 
The high-precision chambers cover a range of $|\eta|< 2.7$. The muon trigger system covers the range $|\eta|< 2.4$ with 
resistive plate chambers in the barrel, and thin gap chambers in the endcap regions.

\section{Event simulation} 
\label{sec:Simulation}

\par MC simulations are used to calculate efficiencies and acceptances for the \Zgll\ signal processes and to unfold the measured \phist\ spectrum for detector effects and for different levels of QED FSR. 
The {\sc{}Powheg} MC generator is used with  {\sc CT10}~\cite{CT10} parton distribution functions (PDFs) to generate both the \Zgee\ and \Zgmm\ signal events. 
It is interfaced to {\sc{}Pythia~6.4} with the {\sc AUET2B-CTEQ6L1} tune~\cite{ATL-PHYS-PUB-2011-009} to simulate the parton shower and the underlying event. 
Generated events are re-weighted as a function of  \ptz\ to the predictions from {\sc ResBos}, which describes the \ptz\ spectrum more accurately ~\cite{Aad:2011gj}. 
Simulated events are also used to estimate background contributions.
The electroweak background processes \Wln\ and 
\Zgtt\ are generated using {\sc Pythia~6.4}. The production of \ttbar\ events is modelled 
using \mcatnlo\ and diboson processes are simulated using {\sc{}Herwig}. The event generators 
are interfaced to {\sc Photos}~\cite{photos} to simulate QED FSR for all of the simulated samples, 
except {\sc Sherpa} which is interfaced to an implementation of the YFS algorithm~\cite{Yennie1961379,Schonherr:2008av}.

Multiple interactions per bunch crossing (pile-up) are accounted for by overlaying simulated minimum bias events. 
To match the observed instantaneous luminosity profile, the simulated events are re-weighted to yield the same distribution 
of the number of interactions per bunch crossing as measured in the data. 
The response of the ATLAS detector to the generated particles is modelled using {\sc Geant4}~\cite{geant4}, and the fully simulated events~\cite{aad:2010wqa} are passed through the same reconstruction chain as the data.
Simulated event samples are corrected for differences with respect to the data in the trigger efficiencies, 
lepton reconstruction and identification efficiencies as well as in energy (momentum) scale and resolution. 
The efficiencies are determined by using a tag-and-probe method similar to the one described in Section $4.3$ of 
Ref.~\cite{WZXsInclusive} based on reconstructed $Z$ and $W$ events, while the energy resolution and scale corrections 
are obtained from a fit to the observed $Z$ boson line shape.

\section{Event reconstruction, selection and background estimation}
\label{sec:Selection}

\par Events recorded during periods with stable beam conditions and passing detector and data-quality requirements are selected. 
At least one primary vertex reconstructed from at least three tracks is required in each event. 
\par Events in the electron channel are selected online by requiring a single electron candidate with a threshold in transverse momentum 
$p_{\rm T}$ that was increased during the data-taking from $20$~\GeV\ to $22$~\GeV\ in response to increased LHC luminosity. 
Electrons are reconstructed from a cluster of cells with significant energy deposits in the electromagnetic calorimeter matched to 
an inner detector track. 
Electron reconstruction uses track refitting with a Gaussian-sum filter to be less sensitive to bremsstrahlung losses and improve the 
estimates of the electron track parameters~\cite{GSF,ATLAS-CONF-2012-047}.
The typical angular resolutions in the electron direction measurements are $0.6$~mrad for $\phi$ and $0.0012$ for $\eta$.
The highest and second highest $p_{\rm T}$ electrons are required to have a transverse momentum $p_{\rm T}^e>25$~\GeV\ and $p_{\rm T}^e>20$~\GeV, 
respectively. 
The electron pseudorapidity must satisfy $|\eta^e|<2.4$ with the calorimeter barrel/endcap transition region $1.37 <|\eta^e|<1.52$ excluded. 
Electrons are required to pass ``medium'' identification criteria based on shower shape and track-quality variables, as described in Refs.~\cite{Aad:2011mk,Aad:2011dm}. 
The criteria are re-optimised for both higher pile-up conditions and higher instantaneous luminosity in $2011$.
\par Events in the muon channel are selected online by a trigger requiring a single muon candidate with $p_{\rm T}^{\mu}>18\GeV$. 
Muons are 
identified as tracks reconstructed in the muon spectrometer matched to tracks reconstructed in the inner detector
and are required to have $p_{\rm T}^{\mu} > 20 \GeV$ and $|\eta^{\mu}| < 2.4$. Only isolated muons are selected by requiring 
the scalar sum of the $p_{\rm T}$ of the tracks within a cone $\Delta R = 0.2$ around the muon to be less than $10\%$ of the muon 
$p_{\rm T}$. 
Muons are required to have a longitudinal impact parameter with respect to the primary vertex less than $10$~mm to reduce
contributions from cosmic-ray muons and in-time pile-up. In addition, the transverse impact parameter of the track 
with respect to the primary vertex divided by its uncertainty must be smaller than ten to reduce non-prompt muon backgrounds.
The typical angular resolutions in the muon direction measurements are $0.4$~mrad for $\phi$ and $0.001$ for $\eta$.

\par \Zgll\ events are selected by requiring two oppositely charged same-flavour leptons with an invariant mass $66\GeV<m_{\ell\ell}<116\GeV$.
After these selection requirements $1.22\cdot10^{6}$ dielectron and $1.69\cdot10^{6}$ dimuon candidate events are found in data.

\par Background contributions from $Z/\gamma^* \rightarrow \tau^+ \tau^-$, $W \to \ell\nu$, $t\bar{t}$ and diboson production
are estimated using MC simulations. The cross sections are normalised to next-to-next-to-leading-order (NNLO) predictions 
for $Z/\gamma^*$ and $W$ production using {\sc{}Fewz}, NLL-NLO predictions for $t\bar{t}$ production~\cite{WZXsInclusive}
and NLO predictions for diboson production~\cite{Campbell:2011bn}.
For both the \eepm\ and \mmpm\ channels, the main background at high $\phist$ values arises from $t\bar{t}$ and diboson production.

\par At low $\phist$ values the background is dominated by multi-jet production, where a jet is falsely identified as a primary $e$ 
or $\mu$. In this case the background is determined by data-driven methods. 
A data event sample dominated by jets faking electrons or muons in the final state is employed to determine the shape of the multi-jet background.
For the \eepm\ channel, the multi-jet sample is obtained from electrons failing the medium identification criteria.
In order to assess systematic uncertainties in the shape of the multi-jet background, an alternative multi-jet control sample was also selected using non-isolated electrons. 
For the \mmpm\ channel, the multi-jet sample is extracted by inverting the isolation requirement on muons.
The uncertainty in its shape was studied by comparing same-sign and opposite-sign dimuon events. 
The normalisation of this multi-jet background template is determined by adjusting the sum of it and other background and signal MC predictions to data as a function of the invariant mass spectrum of the dilepton pair. 
%
\begin{table*}[!hbt]
\caption{The measured normalised differential cross section \normdiffxsfid\
in bins of $\phist$ for $Z/\gamma^* \rightarrow e^+e^-$ and $Z/\gamma^* \rightarrow \mu^+\mu^-$ channels.
The cross sections, which are to be multiplied for convenience by a factor $f$, are reported with respect to the three different treatments of QED final-state radiation. The relative statistical ($\delta_{{\rm stat}}$) and total systematic ($\delta_{{\rm sys}}$) uncertainties are given in percent.
The overall point-to-point uncorrelated additional uncertainty in QED FSR of $0.3\%$ is not included. }
\label{tab:e_mu_total_xsection}
\vspace*{-0.cm}
\begin{center}
\footnotesize{
\scalebox{0.94}{
\begin{tabular}{c|cccccc|cccccc} 
\hline 
 & \multicolumn{6}{l|}{$Z/\gamma^* \rightarrow e^+e^-$} & \multicolumn{6}{l}{$Z/\gamma^* \rightarrow \mu^+\mu^-$} \\ 
 $\phist$  & \multicolumn{4}{l}{$1/\sigma^{{\rm fid}} \cdot {\rm d}\sigma^{{\rm fid}}/{\rm d}\phi^*_\eta$ } & \multicolumn{1}{c}{$\delta_{{\rm stat}}$}& \multicolumn{1}{c|}{$\delta_{{\rm sys}}$} &\multicolumn{4}{l}{$1/\sigma^{{\rm fid}} \cdot {\rm d}\sigma^{{\rm fid}}/{\rm d}\phi^*_\eta$} & \multicolumn{1}{c}{$\delta_{{\rm stat}}$}& \multicolumn{1}{c}{$\delta_{{\rm sys}}$} \\ 
bin range &  Born & dressed & bare & $f$  & [\%] & [\%] & Born & dressed & bare & $f$  & [\%] & [\%] \\ 
\hline 
0.000 -- 0.004 & 9.77 & 9.69 & 9.70 & 1 & 0.46 & 0.35 & 9.77 & 9.67 & 9.67 & 1 & 0.39 & 0.28    \\ 
0.004 -- 0.008 & 9.68 & 9.59 & 9.59 & 1 & 0.47 & 0.26 & 9.76 & 9.66 & 9.66 & 1 & 0.39 & 0.18    \\ 
0.008 -- 0.012 & 9.42 & 9.36 & 9.38 & 1 & 0.47 & 0.28 & 9.42 & 9.34 & 9.35 & 1 & 0.40 & 0.24    \\ 
0.012 -- 0.016 & 9.14 & 9.06 & 9.07 & 1 & 0.48 & 0.35 & 9.26 & 9.17 & 9.18 & 1 & 0.40 & 0.24    \\ 
0.016 -- 0.020 & 8.82 & 8.76 & 8.77 & 1 & 0.49 & 0.24 & 8.83 & 8.76 & 8.77 & 1 & 0.41 & 0.19    \\ 
0.020 -- 0.024 & 8.48 & 8.43 & 8.43 & 1 & 0.50 & 0.25 & 8.51 & 8.44 & 8.45 & 1 & 0.42 & 0.27    \\ 
0.024 -- 0.029 & 7.97 & 7.93 & 7.94 & 1 & 0.46 & 0.26 & 8.05 & 8.00 & 8.01 & 1 & 0.39 & 0.24    \\ 
0.029 -- 0.034 & 7.57 & 7.52 & 7.53 & 1 & 0.47 & 0.22 & 7.57 & 7.52 & 7.53 & 1 & 0.40 & 0.19    \\ 
0.034 -- 0.039 & 7.02 & 7.00 & 7.01 & 1 & 0.49 & 0.29 & 7.11 & 7.09 & 7.09 & 1 & 0.41 & 0.17    \\ 
0.039 -- 0.045 & 6.55 & 6.53 & 6.53 & 1 & 0.46 & 0.22 & 6.50 & 6.49 & 6.49 & 1 & 0.39 & 0.17    \\ 
0.045 -- 0.051 & 5.93 & 5.92 & 5.92 & 1 & 0.48 & 0.22 & 6.00 & 5.99 & 5.99 & 1 & 0.41 & 0.16    \\ 
0.051 -- 0.057 & 5.52 & 5.52 & 5.52 & 1 & 0.50 & 0.22 & 5.52 & 5.53 & 5.53 & 1 & 0.42 & 0.22    \\ 
0.057 -- 0.064 & 5.04 & 5.04 & 5.04 & 1 & 0.48 & 0.22 & 5.00 & 5.01 & 5.01 & 1 & 0.41 & 0.16    \\ 
0.064 -- 0.072 & 4.55 & 4.56 & 4.56 & 1 & 0.48 & 0.22 & 4.52 & 4.52 & 4.52 & 1 & 0.41 & 0.23    \\ 
0.072 -- 0.081 & 4.01 & 4.03 & 4.03 & 1 & 0.48 & 0.21 & 4.04 & 4.06 & 4.06 & 1 & 0.40 & 0.18    \\ 
0.081 -- 0.091 & 3.58 & 3.59 & 3.59 & 1 & 0.48 & 0.22 & 3.53 & 3.55 & 3.55 & 1 & 0.41 & 0.19    \\ 
0.091 -- 0.102 & 3.15 & 3.16 & 3.16 & 1 & 0.49 & 0.23 & 3.14 & 3.16 & 3.16 & 1 & 0.41 & 0.21    \\ 
0.102 -- 0.114 & 2.73 & 2.74 & 2.74 & 1 & 0.50 & 0.26 & 2.73 & 2.75 & 2.74 & 1 & 0.43 & 0.23    \\ 
0.114 -- 0.128 & 2.34 & 2.35 & 2.35 & 1 & 0.50 & 0.29 & 2.35 & 2.37 & 2.37 & 1 & 0.42 & 0.24    \\ 
0.128 -- 0.145 & 2.00 & 2.01 & 2.01 & 1 & 0.49 & 0.24 & 2.00 & 2.01 & 2.01 & 1 & 0.42 & 0.21    \\ 
0.145 -- 0.165 & 1.687 & 1.697 & 1.698 & 1 & 0.49 & 0.28 & 1.669 & 1.680 & 1.679 & 1 & 0.42 & 0.28    \\ 
0.165 -- 0.189 & 1.355 & 1.364 & 1.363 & 1 & 0.50 & 0.25 & 1.355 & 1.365 & 1.365 & 1 & 0.43 & 0.19    \\ 
0.189 -- 0.219 & 1.079 & 1.087 & 1.087 & 1 & 0.50 & 0.23 & 1.086 & 1.093 & 1.093 & 1 & 0.43 & 0.20    \\ 
0.219 -- 0.258 & 8.27 & 8.34 & 8.32 &  10$^{-1}$ & 0.50 & 0.24 & 8.22 & 8.28 & 8.27 &  10$^{-1}$ & 0.43 & 0.19    \\ 
0.258 -- 0.312 & 5.97 & 6.00 & 5.99 &  10$^{-1}$ & 0.50 & 0.25 & 5.94 & 5.97 & 5.97 &  10$^{-1}$ & 0.43 & 0.17    \\ 
0.312 -- 0.391 & 3.97 & 3.99 & 3.99 &  10$^{-1}$ & 0.51 & 0.22 & 3.94 & 3.96 & 3.96 &  10$^{-1}$ & 0.44 & 0.17    \\ 
0.391 -- 0.524 & 2.28 & 2.28 & 2.28 &  10$^{-1}$ & 0.52 & 0.24 & 2.29 & 2.29 & 2.29 &  10$^{-1}$ & 0.45 & 0.19    \\ 
0.524 -- 0.695 & 1.176 & 1.179 & 1.177 & 10$^{-1}$ & 0.64 & 0.29 & 1.164 & 1.168 & 1.166 & 10$^{-1}$ & 0.55 & 0.23    \\ 
0.695 -- 0.918 & 5.79 & 5.80 & 5.79 &  10$^{-2}$ & 0.79 & 0.37 & 5.77 & 5.79 & 5.78 &  10$^{-2}$ & 0.69 & 0.29    \\ 
0.918 -- 1.153 & 2.94 & 2.95 & 2.95 &  10$^{-2}$ & 1.07 & 0.47 & 2.91 & 2.91 & 2.91 &  10$^{-2}$ & 0.95 & 0.37    \\ 
1.153 -- 1.496 & 1.54 & 1.55 & 1.54 &  10$^{-2}$ & 1.22 & 0.52 & 1.50 & 1.51 & 1.51 &  10$^{-2}$ & 1.09 & 0.41    \\ 
1.496 -- 1.947 & 7.25 & 7.26 & 7.25 &  10$^{-3}$ & 1.55 & 0.66 & 7.05 & 7.06 & 7.06 &  10$^{-3}$ & 1.39 & 0.48    \\ 
1.947 -- 2.522 & 3.52 & 3.51 & 3.50 &  10$^{-3}$ & 1.97 & 0.78 & 3.56 & 3.56 & 3.55 &  10$^{-3}$ & 1.74 & 0.62    \\ 
2.522 -- 3.277 & 1.73 & 1.73 & 1.72 &  10$^{-3}$ & 2.46 & 0.96 & 1.79 & 1.80 & 1.80 &  10$^{-3}$ & 2.13 & 0.71    \\ 
\hline 
\end{tabular} 
}

}
\end{center}
\vspace*{-0.5cm}
\end{table*}
%
An extended dilepton mass range, $50\GeV<m_{\ell\ell}<150\GeV$ ($200$~GeV for electrons), was 
employed to better constrain the off-resonance region and improve the accuracy of the multi-jet background normalisation.

\par The total fraction of background 
events is $(0.61\pm 0.31)\%$ in the \eepm\ channel and $(0.56\pm 0.28)\%$ in the \mmpm\ channel.
The multi-jet background represents  $\sim 50\%$ of the total background in both channels and dominates at low $\phist$ values.
An irreducible background may also arise from the production of a lepton pair via photon-photon interactions, $\gamma \gamma \rightarrow \ell^+ \ell^-$.
This contribution was evaluated at leading order using FEWZ~3.1~\cite{Melnikov:2006kv,Li:2012wn} and the MRST2004qed~\cite{Martin:2004dh} PDF, currently the only available PDF set containing a description of the QED part of the proton.
According to the LO cross section calculated in the fiducial lepton acceptance, the fraction of photon-induced events is expected to be below $0.1\%$, with an uncertainty of $50\%$.
This contribution is six times lower than the sum of other background contributions and is therefore neglected.
%

\section{Cross-section measurement and systematic uncertainties}
\label{sec:Measurement}

The differential cross section is evaluated in bins of $\phist$, or of $(\phist, y_Z)$, from the number of observed data events in each bin after subtraction of the estimated number of background events.

\par A bin-by-bin correction is used to correct the observed data for detector acceptances and inefficiencies, as well as for QED FSR.
The correction factors are determined using signal MC events.
For the chosen bin widths the purity, defined as the fraction of simulated events reconstructed in a $\phist$ bin which have generator-level $\phist$ in the same bin, 
is always more than $83\%$ and reaches $98\%$ in the highest $\phist$ bins.
In each bin, the data are normalised to the cross section integrated over the fiducial acceptance region.

\par An analysis of systematic uncertainties was performed, in which the sensitivity of the measurements to variations in the efficiencies 
and energy scales of the detector components and to the details of the correction procedure is tested. %
The systematic uncertainties in the measured cross section are determined by repeating the
analysis after applying appropriate variations for each source of systematic uncertainty to the simulated samples.
The systematic uncertainties which are correlated between $\phist$ bins are listed below.

\begin{itemize}

\item Uncertainties in the estimation of the number of background events from multi-jet, \Wln\ and \Zgtt\ decays, $t\bar{t}$ and diboson processes 
yield values of up to $0.3\%$ in the \eepm\ and \mmpm\ channels, when propagated to the normalised differential cross section.

\item Possible mis-modelling of the angular resolution of tracking detectors leads to uncertainties of up to $0.3\%$ ($0.2\%$) on the normalised 
differential cross section in the \eepm\ (\mmpm\ ) channel.

\item The dependence of the bin-by-bin correction factors on the shape of the assumed $\phist$ distribution was tested by
re-weighting simulated events to the measured \phist\ cross section. An iterative Bayesian unfolding technique~\cite{D'Agostini:1994zf} 
was employed as an alternative approach
to assess systematic uncertainties. The uncertainty in the correction procedure is found to be smaller than $0.1\%$ in both channels and for 
the full \phist\ range.

\item As the definition of the $\phist$ variable is based on the lepton angles, the normalised differential cross section depends only weakly on uncertainties in the lepton energy/momentum scale and resolution.
When propagated to the normalised differential cross section, these uncertainties amount to less than  $0.1\%$ and $0.03\%$ in the \eepm\ and \mmpm\ channels, respectively.

\item Uncertainties arising from the mis-modelling of lepton identification efficiencies and trigger efficiencies in the simulation 
amount respectively to $0.05\%$ ($0.03\%$) and $0.04\%$ ($0.02\%$) in the \eepm\ (\mmpm) channel.

\item Pile-up has only a weak influence on this measurement and results in an uncertainty of at most $0.05\%$ on the normalised differential cross section.
\end{itemize}

\par A second class of systematic uncertainties, listed below, are considered uncorrelated across \phist\ bins.

\begin{itemize}
\item Uncertainties on the bin-by-bin correction factors arising from the MC sample statistics are $0.2\%$ ($0.13\%$) at low \phist\ in the $\eepm$ ($\mmpm$) channel, increasing to $0.9\%$ ($0.6\%$) in the highest $\phist$ bins.
\item Possible local biases in angular measurements ($\phi$, $\eta$) by tracking detectors yield an estimated constant uncertainty of $0.1\%$ on the normalised differential cross section.
The local effect of these biases allows bin-to-bin correlations to be neglected.
The impact of this assumption on the combination of electron and muon channel results is small.
\item A conservative systematic uncertainty of $0.3\%$ due to $\phist$-dependent modelling of QED FSR is assigned by comparing predictions from {\sc Photos}~\cite{photos} and from the {\sc Sherpa} implementation of the YFS algorithm~\cite{Yennie1961379,Schonherr:2008av}.
This comparison provides the size of the uncertainty but however does not allow the shape of the $\phist$ dependence to be estimated.
This uncertainty was therefore treated as uncorrelated across $\phist$ bins.
The uncertainty is assumed to hold for cross
sections at Born, dressed and bare levels and for both electron and muon
channel measurements. 
It therefore does not affect the combination of them.
\end{itemize}

\par The total systematic uncertainty on each data point is formed by adding the individual contributions in quadrature.
%
\begin{figure}[!hbt] 
\vspace*{-0.3cm}
\begin{center}
\includegraphics[width=0.5\textwidth]{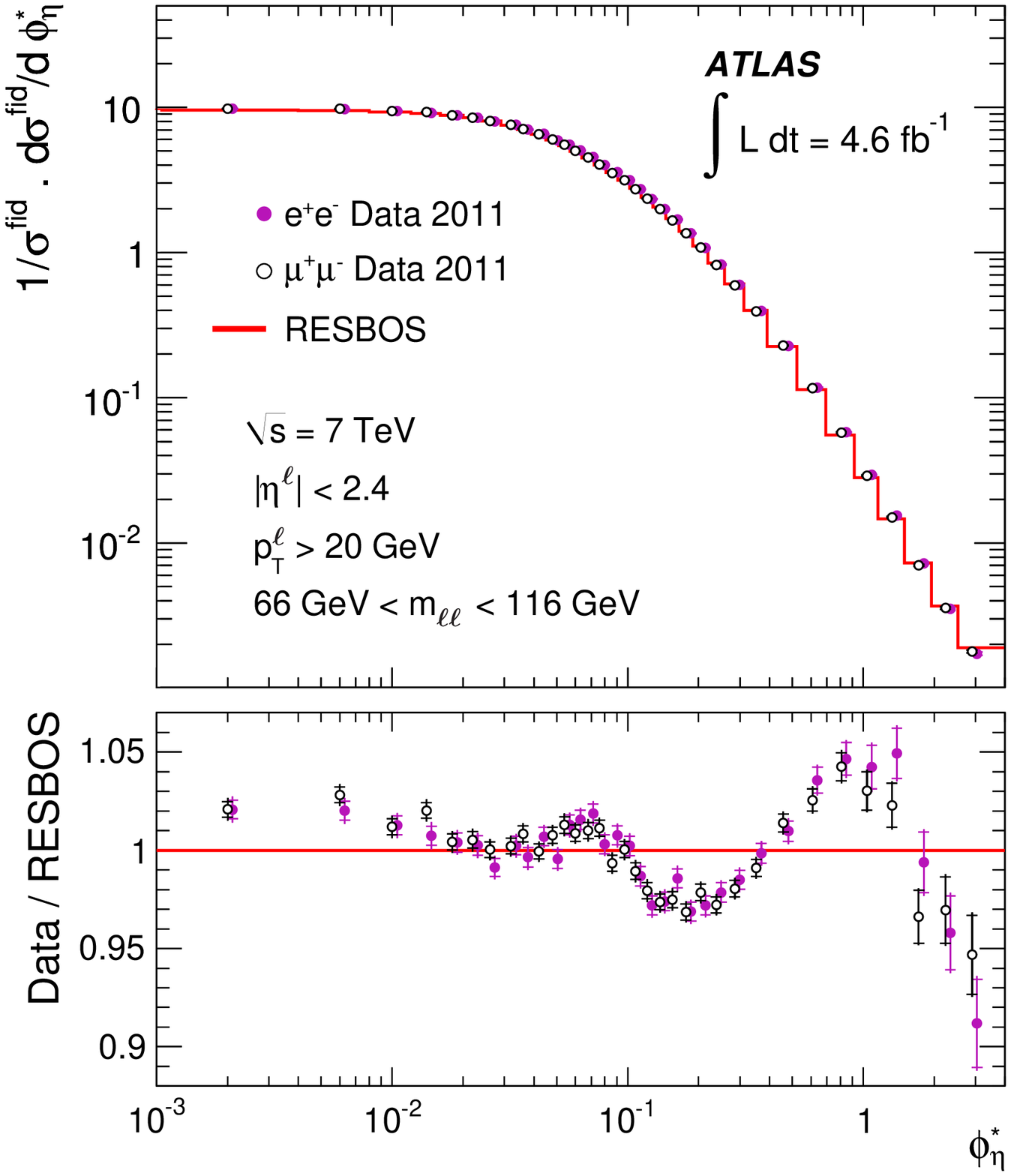}
\end{center}      
\vspace*{-0.5cm}
 \caption{The measured normalised differential cross section \normdiffxsfid\
as a function of $\phist$ for $Z/\gamma^* \rightarrow e^+e^-$ (closed dots) and $Z/\gamma^* \rightarrow \mu^+\mu^-$ (open dots) channels. 
The measurements are compared to {\sc ResBos} predictions represented by a line. 
The ratio of measured cross sections to {\sc ResBos} predictions is presented in the bottom panel. 
The measurements are displaced horizontally for better visibility.
The inner and outer error bars on the data points represent the statistical and total uncertainties, respectively. The uncertainty due to QED FSR is included in the total uncertainties.}
\label{fig:sigma_tot_e_mu}  
\end{figure}
%

%
\section{Results and discussion}
\label{sec:Results}

The normalised differential cross sections measured for $Z/\gamma^* \rightarrow e^+e^-$ and $Z/\gamma^* \rightarrow \mu^+\mu^-$ 
production in the fiducial acceptance are presented in Table~\ref{tab:e_mu_total_xsection}.
The measurements are reported with respect to the Born, dressed and bare reference points at particle level regarding QED FSR.
The QED FSR corrections for the three levels are calculated using {\sc Photos}. 
The measured cross sections defined at the $Z/\gamma^*$ Born level are shown in Fig.~\ref{fig:sigma_tot_e_mu} for the \eepm\ and \mmpm\ channels and are compared to predictions from {\sc ResBos}. 

The normalised differential cross sections measured in the fiducial acceptance for the two channels are combined using a $\chi^2$ minimisation method which takes into account the point-to-point correlated and uncorrelated systematic uncertainties~\cite{Glazov:2005rn,Aaron:2009bp,Aaron:2009wt} and correlations between electron and
muon channels.
The procedure allows a model independent check of the electron and muon data consistency and leads to a significant reduction of the correlated
uncertainties.

%
%
\begin{table}[!hbt]
\caption{The combined normalised differential cross section \normdiffxsfid\
  in bins of $\phist$ at Born level. The statistical ($\delta_{{\rm stat}}$) 
  and total systematic ($\delta_{{\rm sys}}$) uncertainties are given in percent. 
  The normalised differential cross section extrapolated to the full lepton 
  acceptance \normdiffxstot\ is obtained at Born level by multiplication with 
  the inverse acceptance correction factor $A_c^{-1}$. The uncertainty 
  $\delta{(A_c^{-1})}$ on this acceptance correction factor is also given in 
  percent. The overall point-to-point uncorrelated additional uncertainty in 
  QED FSR of $0.3\%$ is not included. }
\label{tab:comb_total_xsection}
\begin{center}
\footnotesize{
\scalebox{0.94}{
\begin{tabular}{c|c|cc|cc} 
\hline 
$\phi^*_\eta$ &  $1/\sigma^{{\rm fid}} \cdot {\rm d}\sigma^{{\rm fid}}/{\rm d}\phi^*_\eta$ & $\delta_{{\rm stat}}$ & $\delta_{{\rm sys}}$ & $A_c^{-1}$ & $\delta{(A_c^{-1})}$ \\ 
bin range &   & $[\%]$ & $[\%]$ & & $[\%]$ \\ 
\hline 
0.000 -- 0.004 & 9.77  & 0.30 & 0.21 & 1.06 & 3.8  \\ 
0.004 -- 0.008 & 9.73  & 0.30 & 0.20 & 1.06 & 3.0  \\ 
0.008 -- 0.012 & 9.41  & 0.31 & 0.18 & 1.06 & 3.7  \\ 
0.012 -- 0.016 & 9.21  & 0.31 & 0.22 & 1.06 & 2.4  \\ 
0.016 -- 0.020 & 8.82  & 0.31 & 0.16 & 1.05 & 2.5  \\ 
0.020 -- 0.024 & 8.49  & 0.32 & 0.18 & 1.05 & 2.2  \\ 
0.024 -- 0.029 & 8.01  & 0.29 & 0.18 & 1.05 & 1.8  \\ 
0.029 -- 0.034 & 7.56  & 0.30 & 0.14 & 1.04 & 2.4  \\ 
0.034 -- 0.039 & 7.07  & 0.31 & 0.15 & 1.04 & 2.2  \\ 
0.039 -- 0.045 & 6.52  & 0.30 & 0.14 & 1.03 & 2.2  \\ 
0.045 -- 0.051 & 5.97  & 0.31 & 0.13 & 1.02 & 2.8  \\ 
0.051 -- 0.057 & 5.52  & 0.32 & 0.16 & 1.01 & 2.1  \\ 
0.057 -- 0.064 & 5.02  & 0.31 & 0.13 & 1.01 & 1.9  \\ 
0.064 -- 0.072 & 4.54  & 0.31 & 0.18 & 1.00 & 2.0  \\ 
0.072 -- 0.081 & 4.03  & 0.31 & 0.13 & 0.99 & 1.8  \\ 
0.081 -- 0.091 & 3.56  & 0.31 & 0.15 & 0.99 & 1.0  \\ 
0.091 -- 0.102 & 3.15  & 0.32 & 0.16 & 0.98 & 1.1  \\ 
0.102 -- 0.114 & 2.731 & 0.32 & 0.17 & 0.97 & 1.3  \\ 
0.114 -- 0.128 & 2.347 & 0.32 & 0.19 & 0.97 & 1.3  \\ 
0.128 -- 0.145 & 1.996 & 0.32 & 0.16 & 0.96 & 1.7  \\ 
0.145 -- 0.165 & 1.677 & 0.32 & 0.19 & 0.95 & 2.0  \\ 
0.165 -- 0.189 & 1.355 & 0.32 & 0.16 & 0.95 & 2.7  \\ 
0.189 -- 0.219 & 1.084 & 0.32 & 0.15 & 0.94 & 2.3  \\ 
0.219 -- 0.258 & 8.24 $\cdot$ 10$^{-1}$ & 0.33 & 0.15 & 0.94 & 2.9  \\ 
0.258 -- 0.312 & 5.95 $\cdot$ 10$^{-1}$ & 0.33 & 0.14 & 0.93 & 2.9  \\ 
0.312 -- 0.391 & 3.96 $\cdot$ 10$^{-1}$ & 0.33 & 0.14 & 0.92 & 3.4  \\ 
0.391 -- 0.524 & 2.282 $\cdot$ 10$^{-1}$ & 0.34 & 0.15 & 0.92 & 3.5  \\ 
0.524 -- 0.695 & 1.169 $\cdot$ 10$^{-1}$ & 0.42 & 0.18 & 0.92 & 4.4  \\ 
0.695 -- 0.918 & 5.78 $\cdot$ 10$^{-2}$ & 0.52 & 0.23 & 0.93 & 4.0  \\ 
0.918 -- 1.153 & 2.92 $\cdot$ 10$^{-2}$ & 0.71 & 0.29 & 0.94 & 5.3  \\ 
1.153 -- 1.496 & 1.52 $\cdot$ 10$^{-2}$ & 0.81 & 0.33 & 0.98 & 10.5  \\ 
1.496 -- 1.947 & 7.13 $\cdot$ 10$^{-3}$ & 1.04 & 0.40 & 1.04 & 10.3  \\ 
1.947 -- 2.522 & 3.54 $\cdot$ 10$^{-3}$ & 1.30 & 0.49 & 1.11 & 17.5  \\ 
2.522 -- 3.277 & 1.77 $\cdot$ 10$^{-3}$ & 1.61 & 0.58 & 1.19 & 16.2  \\ 
\hline 
\end{tabular} 
}

}
\end{center}
\vspace*{-0.5cm}
\end{table}
%
%
The uncertainties due to the unfolding procedure, the pile-up, and QED FSR are considered to be completely correlated between the \eepm\ and \mmpm\ channels.
The minimisation yields a total $\chi^2$ per degree of freedom ($n_{{\rm dof}}$) of $\chi^2/n_{{\rm dof}} = 33.2/34$, indicating a good consistency between the electron and muon data.
Measured values of the combined normalised differential cross section \normdiffxsfid\ 
within the fiducial lepton acceptance are presented in Table~\ref{tab:comb_total_xsection}.
At lower $\phist$ values the statistical and systematic uncertainties are 
of the same order, whilst for large $\phist$ values statistical uncertainties 
are dominating.
The acceptance correction factors $A_c$ needed to extrapolate the measurement to the full lepton acceptance are determined using the {\sc{}Powheg} simulation with the {\sc CT10} PDF set and reweighted as a function of $\zpt$ to {\sc ResBos} predictions.
The uncertainty in $A_c$ is estimated from the extreme differences among predictions obtained with {\sc ResBos}, \mcatnlo, {\sc Sherpa}, {\sc Alpgen}, {\sc{}Herwig} and {\sc Powheg} interfaced to {\sc{}Pythia8}. 
Uncertainties in $A_c$ resulting from PDF uncertainties are below $1\%$.
%
\begin{figure}[!hbt] 
\begin{center}
\includegraphics[width=0.5\textwidth]{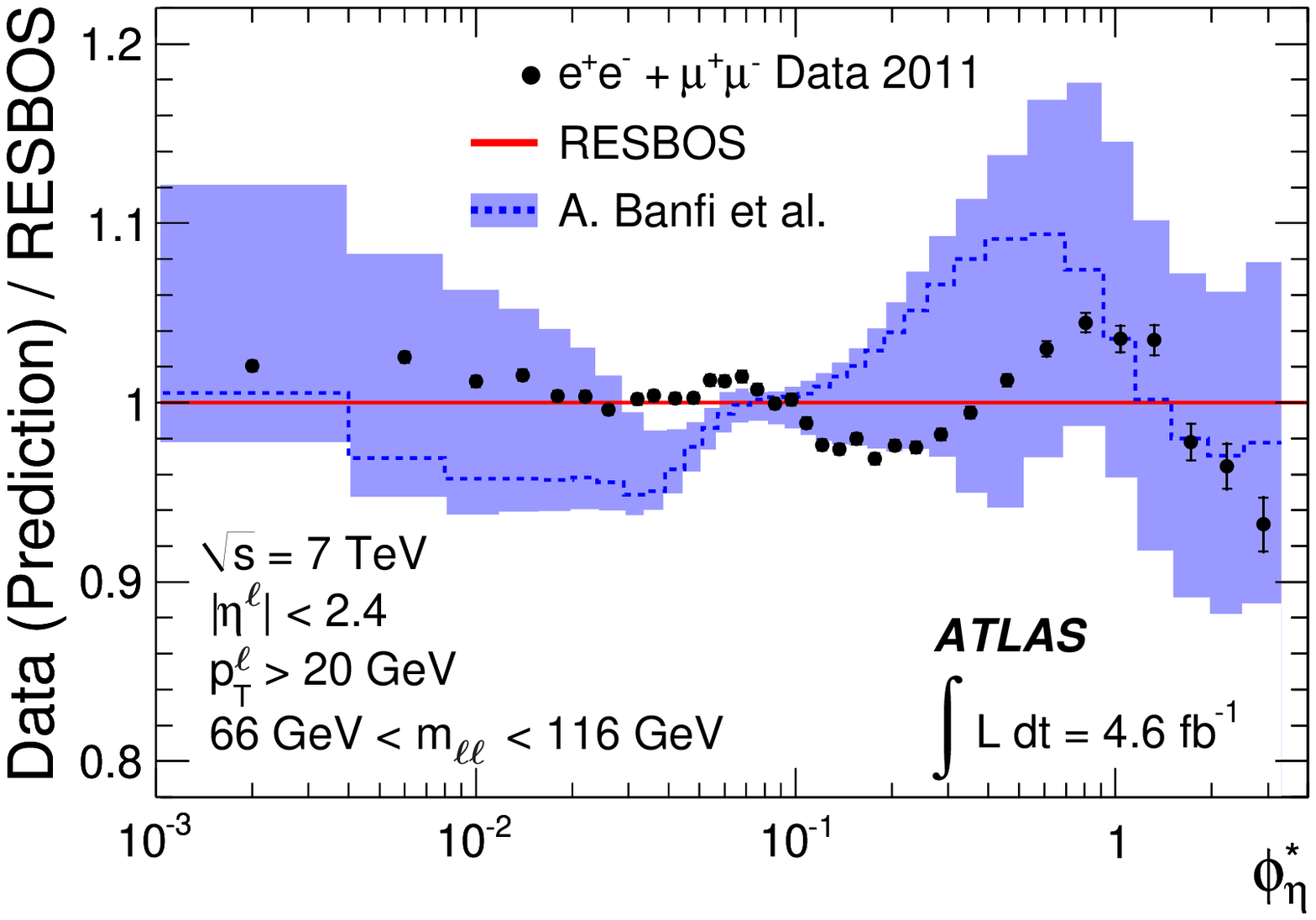}\\
\includegraphics[width=0.5\textwidth]{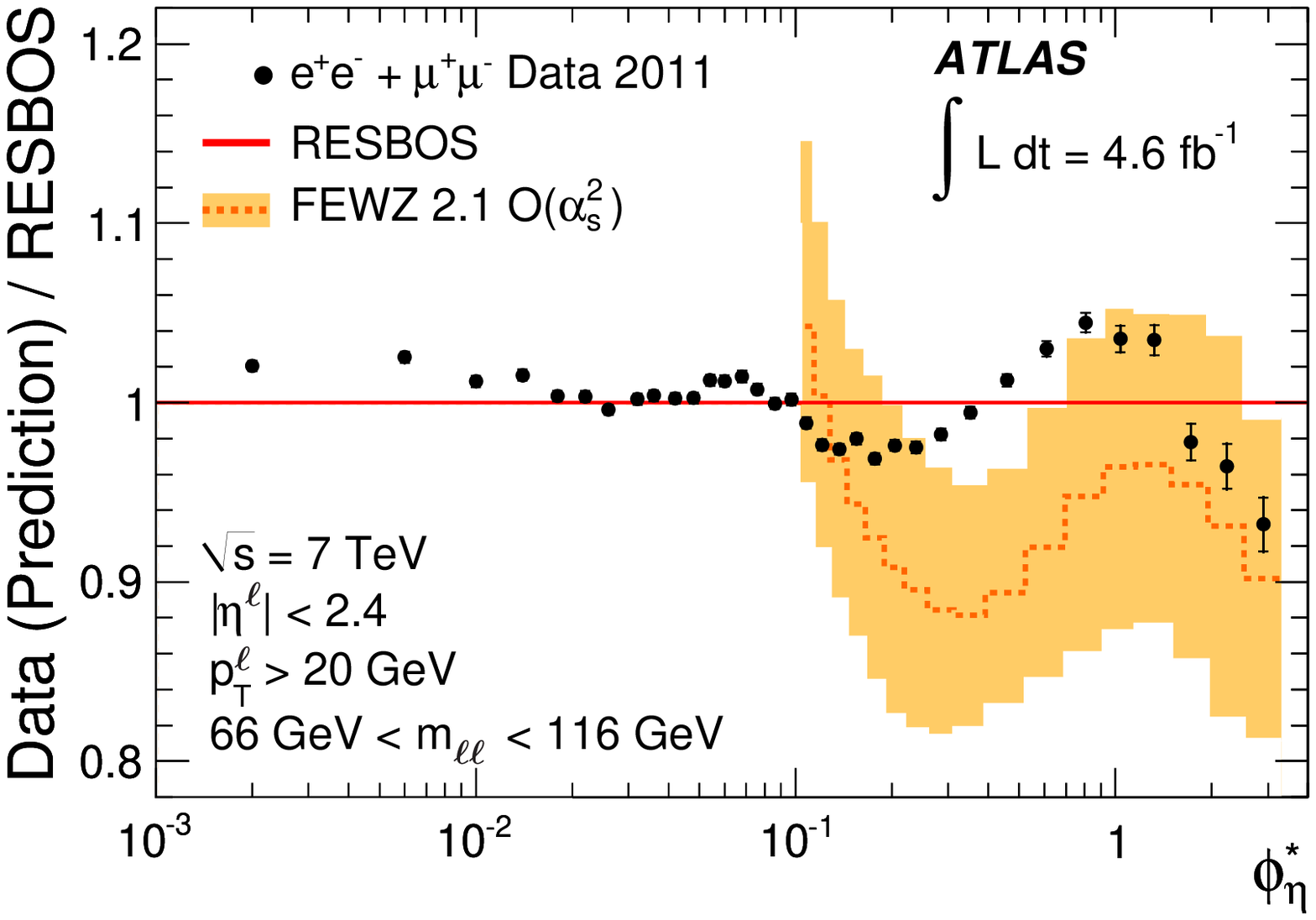}
\end{center}      
\vspace*{-0.7cm}
 \caption{The ratio of the combined normalised differential cross section 
\normdiffxsfid\ to {\sc ResBos} predictions as a function of $\phist$. 
The inner and outer error bars on the data points represent the statistical and total uncertainties, respectively. The uncertainty due to
QED FSR is included in the total uncertainties. The measurements are also compared to predictions, which are represented by a dashed line, from Ref.~\protect{\cite{Banfi:2012du}} and from {\sc Fewz} in the top and bottom panels, respectively. Uncertainties associated with these two calculations are represented by shaded bands.
The prediction from {\sc Fewz} is only presented for $\phist > 0.1$. }
\label{fig:sigma_theory_comp}  
\end{figure}
%
%

\par The ratio of the combined normalised differential cross section to the {\sc ResBos} prediction is shown as a function of $\phist$ in Fig.~\ref{fig:sigma_theory_comp}.
The measurement is also compared to a QCD calculation by A.~Banfi {\it et al.}~\cite{Banfi:2012du} and to another obtained with {\sc Fewz}~2.1. 
The ratios of these two calculations to {\sc ResBos} predictions are also shown in Fig.~\ref{fig:sigma_theory_comp}.
The CTEQ6m~\cite{Pumplin:2002vw} PDF set is used in the calculations in Ref.~\cite{Banfi:2012du}. 
The theoretical uncertainties on this calculation are evaluated by varying the resummation, renormalisation and factorisation scales $\mu_Q$, $\mu_R$ and $\mu_F$ between $m_Z/2$ and $2m_Z$, with the constraints $0.5 \leq \mu_i/\mu_j \leq 2$, where $i, j \in \{F,Q,R \}$, and $\mu_F/\mu_Q \geq 1$.
Uncertainties coming from the PDFs are also considered~\cite{Banfi:2012du}.
For {\sc Fewz}, the {\sc CT10} PDF set is used.
Uncertainties are evaluated by varying $\mu_R$ and $\mu_F$ by factors of two around the nominal scale $m_Z$ with the constraint $0.5 \leq \mu_R / \mu_F \leq 2$, by varying $\alpha_s$ within a range corresponding to $90\%$ confidence-level (CL) limits~\cite{Martin:2009bu}, and by using the PDF error eigenvector sets.
%
%
%
\begin{table*}[hbt]
\caption{The combined normalised differential cross section 
\normdiffxsfid\ in bins of $\phist$ and in three $|y_Z|$ ranges. 
The statistical ($\delta_{{\rm stat}}$) and total systematic ($\delta_{{\rm sys}}$) uncertainties are given in percent.
The overall point-to-point uncorrelated additional uncertainty in QED FSR of $0.3\%$ is not included.}
\label{tab:comb_ybins_xsection}
\begin{center}
\footnotesize{
\scalebox{0.94}{
\begin{tabular}{c|ccc|ccc|ccc} 
\hline 
 & \multicolumn{3}{c}{ $|y_Z| < 0.8$} &  \multicolumn{3}{|c|}{ $0.8 \leq |y_Z| < 1.6$} &  \multicolumn{3}{c}{ $|y_Z| \geq 1.6$} \\ 
$\phi^*_\eta$ &  $1/\sigma^{{\rm fid}} \cdot {\rm d}\sigma^{{\rm fid}}/{\rm d}\phi^*_\eta$ & $\delta_{{\rm stat}}$ &$\delta_{{\rm sys}}$ &  $1/\sigma^{{\rm fid}} \cdot {\rm d}\sigma^{{\rm fid}}/{\rm d}\phi^*_\eta$ & $\delta_{{\rm stat}}$ &$\delta_{{\rm sys}}$&  $1/\sigma^{{\rm fid}} \cdot {\rm d}\sigma^{{\rm fid}}/{\rm d}\phi^*_\eta$ & $\delta_{{\rm stat}}$ &$\delta_{{\rm sys}}$\\ 
bin range &   & $[\%]$ & $[\%]$ &   & $[\%]$ & $[\%]$ &   & $[\%]$ & $[\%]$  \\ 
\hline 
0.000 -- 0.004 & 9.73  & 0.45 & 0.25 & 9.81  & 0.48 & 0.21 & 9.79  & 0.75 & 0.33 \\ 
0.004 -- 0.008 & 9.65  & 0.45 & 0.23 & 9.81  & 0.48 & 0.26 & 9.77  & 0.75 & 0.35 \\ 
0.008 -- 0.012 & 9.37  & 0.46 & 0.24 & 9.45  & 0.49 & 0.23 & 9.45  & 0.76 & 0.30 \\ 
0.012 -- 0.016 & 9.12  & 0.46 & 0.24 & 9.31  & 0.49 & 0.25 & 9.25  & 0.77 & 0.38 \\ 
0.016 -- 0.020 & 8.81  & 0.47 & 0.21 & 8.88  & 0.50 & 0.22 & 8.72  & 0.79 & 0.31 \\ 
0.020 -- 0.024 & 8.49  & 0.48 & 0.20 & 8.48  & 0.51 & 0.25 & 8.52  & 0.80 & 0.38 \\ 
0.024 -- 0.029 & 7.99  & 0.44 & 0.23 & 8.05  & 0.47 & 0.21 & 7.99  & 0.74 & 0.29 \\ 
0.029 -- 0.034 & 7.54  & 0.46 & 0.21 & 7.64  & 0.48 & 0.19 & 7.45  & 0.77 & 0.28 \\ 
0.034 -- 0.039 & 7.12  & 0.47 & 0.19 & 7.07  & 0.50 & 0.21 & 6.99  & 0.79 & 0.31 \\ 
0.039 -- 0.045 & 6.53  & 0.45 & 0.18 & 6.57  & 0.47 & 0.19 & 6.38  & 0.75 & 0.27 \\ 
0.045 -- 0.051 & 5.92  & 0.47 & 0.18 & 6.01  & 0.50 & 0.19 & 6.02  & 0.78 & 0.28 \\ 
0.051 -- 0.057 & 5.52  & 0.48 & 0.20 & 5.50  & 0.52 & 0.21 & 5.53  & 0.81 & 0.30 \\ 
0.057 -- 0.064 & 5.06  & 0.47 & 0.18 & 5.00  & 0.50 & 0.19 & 4.97  & 0.79 & 0.28 \\ 
0.064 -- 0.072 & 4.53  & 0.46 & 0.22 & 4.53  & 0.49 & 0.21 & 4.56  & 0.77 & 0.30 \\ 
0.072 -- 0.081 & 4.02  & 0.46 & 0.19 & 4.04  & 0.49 & 0.18 & 4.02  & 0.77 & 0.27 \\ 
0.081 -- 0.091 & 3.56  & 0.47 & 0.19 & 3.55  & 0.50 & 0.20 & 3.55  & 0.78 & 0.27 \\ 
0.091 -- 0.102 & 3.15  & 0.47 & 0.20 & 3.14  & 0.50 & 0.21 & 3.15  & 0.79 & 0.28 \\ 
0.102 -- 0.114 & 2.72  & 0.49 & 0.20 & 2.73  & 0.52 & 0.21 & 2.75  & 0.81 & 0.29 \\ 
0.114 -- 0.128 & 2.34  & 0.48 & 0.22 & 2.34  & 0.52 & 0.22 & 2.37  & 0.81 & 0.31 \\ 
0.128 -- 0.145 & 2.00  & 0.47 & 0.19 & 2.00  & 0.51 & 0.22 & 1.99  & 0.80 & 0.27 \\ 
0.145 -- 0.165 & 1.667 & 0.48 & 0.20 & 1.677 & 0.51 & 0.21 & 1.707 & 0.80 & 0.38 \\ 
0.165 -- 0.189 & 1.342 & 0.49 & 0.19 & 1.356 & 0.52 & 0.21 & 1.385 & 0.81 & 0.28 \\ 
0.189 -- 0.219 & 1.073 & 0.49 & 0.19 & 1.085 & 0.52 & 0.19 & 1.112 & 0.81 & 0.28 \\ 
0.219 -- 0.258 & 8.18 $\cdot$ 10$^{-1}$ & 0.49 & 0.18& 8.22 $\cdot$ 10$^{-1}$ & 0.52 & 0.19& 8.47 $\cdot$ 10$^{-1}$ & 0.81 & 0.28   \\ 
0.258 -- 0.312 & 5.96 $\cdot$ 10$^{-1}$ & 0.49 & 0.18& 5.87 $\cdot$ 10$^{-1}$ & 0.53 & 0.19& 6.13 $\cdot$ 10$^{-1}$ & 0.81 & 0.26   \\ 
0.312 -- 0.391 & 3.95 $\cdot$ 10$^{-1}$ & 0.50 & 0.18& 3.89 $\cdot$ 10$^{-1}$ & 0.54 & 0.19& 4.14 $\cdot$ 10$^{-1}$ & 0.82 & 0.26   \\ 
0.391 -- 0.524 & 2.28 $\cdot$ 10$^{-1}$ & 0.50 & 0.20& 2.25 $\cdot$ 10$^{-1}$ & 0.55 & 0.20& 2.36 $\cdot$ 10$^{-1}$ & 0.84 & 0.28   \\ 
0.524 -- 0.695 & 1.174 $\cdot$ 10$^{-1}$ & 0.62 & 0.24 & 1.151 $\cdot$ 10$^{-1}$ & 0.67 & 0.25& 1.21 $\cdot$ 10$^{-1}$ & 1.03 & 0.34 \\ 
0.695 -- 0.918 & 5.88 $\cdot$ 10$^{-2}$ & 0.77 & 0.30& 5.70 $\cdot$ 10$^{-2}$ & 0.84 & 0.31& 5.69 $\cdot$ 10$^{-2}$ & 1.32 & 0.42   \\ 
0.918 -- 1.153 & 3.05 $\cdot$ 10$^{-2}$ & 1.04 & 0.40& 2.87 $\cdot$ 10$^{-2}$ & 1.15 & 0.42& 2.67 $\cdot$ 10$^{-2}$ & 1.86 & 0.58   \\ 
1.153 -- 1.496 & 1.62 $\cdot$ 10$^{-2}$ & 1.17 & 0.44& 1.50 $\cdot$ 10$^{-2}$ & 1.30 & 0.47& 1.29 $\cdot$ 10$^{-2}$ & 2.21 & 0.71   \\ 
1.496 -- 1.947 & 7.67 $\cdot$ 10$^{-3}$ & 1.50 & 0.56& 7.08 $\cdot$ 10$^{-3}$ & 1.65 & 0.58& 5.66 $\cdot$ 10$^{-3}$ & 2.91 & 0.95   \\ 
1.947 -- 2.522 & 4.08 $\cdot$ 10$^{-3}$ & 1.83 & 0.70& 3.53 $\cdot$ 10$^{-3}$ & 2.06 & 0.69& 2.00 $\cdot$ 10$^{-3}$ & 4.35 & 1.39   \\ 
2.522 -- 3.277 & 2.10 $\cdot$ 10$^{-3}$ & 2.22 & 0.80& 1.70 $\cdot$ 10$^{-3}$ & 2.59 & 0.87& 9.66 $\cdot$ 10$^{-4}$ & 5.39 & 1.92   \\ 
\hline 
\end{tabular} 
}

}
\end{center}
\vspace*{-0.5cm}
\end{table*}

\begin{figure*}[!hbt] 
\begin{center}
\includegraphics[width=0.96\textwidth]{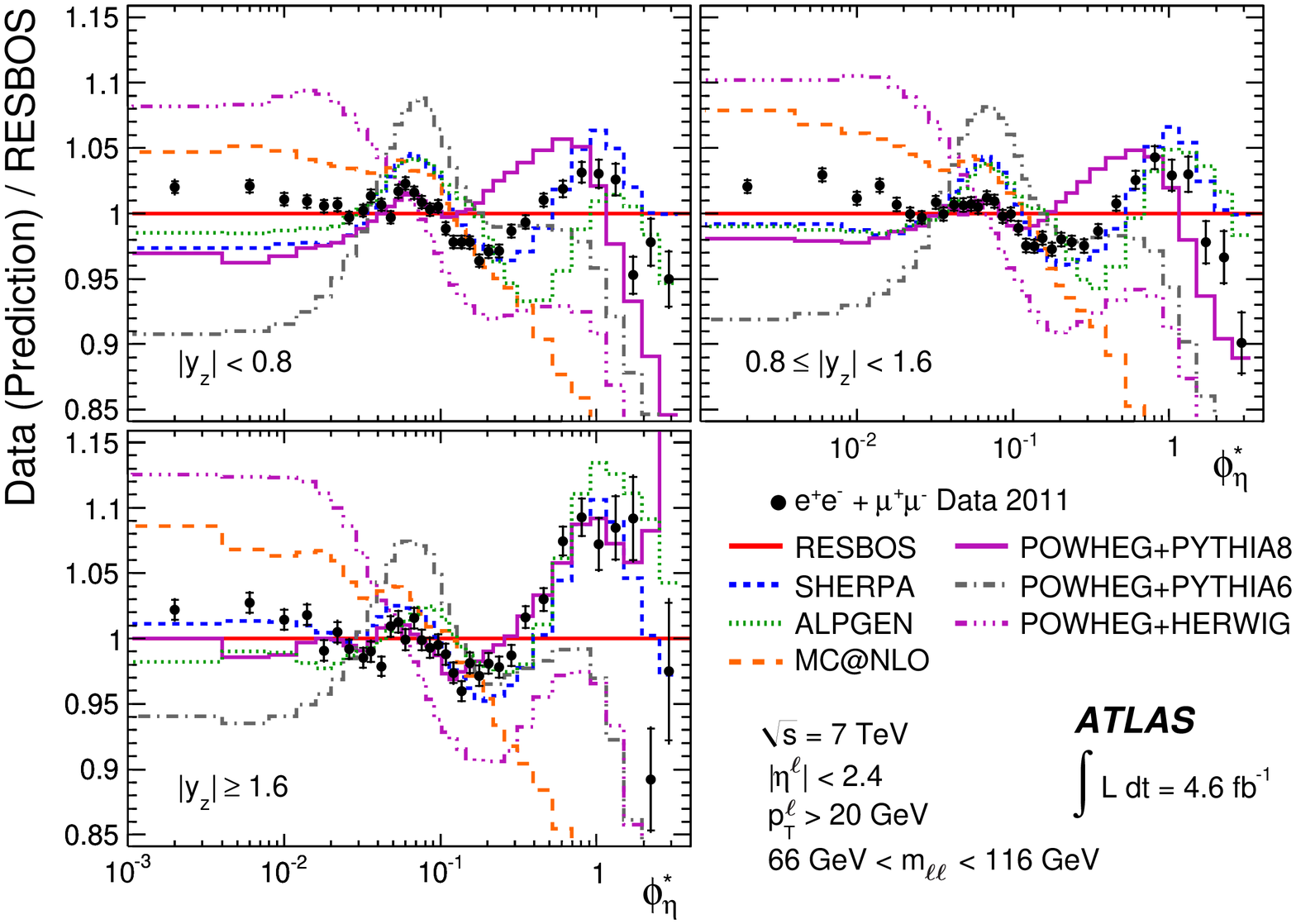}
\end{center}      
\vspace*{-0.7cm}
 \caption{The ratio of the combined normalised differential cross section \normdiffxsfid\ to the {\sc ResBos} predictions 
   as a function of $\phist$ in three ranges of $|y_Z|$. The inner and outer error bars on the data points represent the 
   statistical and total uncertainties, respectively. The uncertainty due to QED FSR is included in the total uncertainties. 
   The measurements are also compared to predictions from different MC event generators.}
\label{fig:sigma_ybins}  
\vspace*{-0.5cm}
\end{figure*}

\par The difference between the {\sc ResBos} prediction and data is \mbox{$\sim 2\%$} for $\phist < 0.1$, increasing to $5\%$ for higher $\phist$ values.
This difference is smaller than the uncertainty in {\sc ResBos} predictions due to the propagation of PDF eigenvectors sets, which amounts to $4\%$ for $\phist < 0.1$ and $6\%$ above.
The description of data provided by calculations from A.~Banfi {\it et al.}~\cite{Banfi:2012du} is less good than {\sc ResBos} but observed differences remain within the 
theoretical uncertainties of the calculation.
The prediction obtained with {\sc Fewz} undershoots the data by $\sim$~$10\%$, as already observed for the \zpt\ spectrum in Ref.~\cite{Aad:2011gj}.
At low $\phist$ values, corresponding mainly to low \ptz, fixed-order perturbative QCD calculations are not expected to give an 
adequate description of the cross section. 
The prediction from {\sc Fewz} is therefore only presented for $\phist > 0.1$. 
It is normalised using the total cross section predicted by {\sc Fewz}, which accurately describes experimental measurements~\cite{Aad:2011dm}.

The cross section is also measured double differentially in bins of $\phist$ for three independent bins of $|y_Z|$ for both the \eepm\ and \mmpm\ channels.
The double differential cross-section measurements in the two channels are combined using the same $\chi^2$ minimisation procedure as used for the single differential cross section.
The minimisation yields a total $\chi^2/n_{{\rm dof}} = 118/102$.
Measured values of the combined normalised differential cross section \normdiffxsfid\ 
within the fiducial lepton acceptance in all $\phist$  and $|y_Z|$ bins are presented in Table~\ref{tab:comb_ybins_xsection}.

\par The ratio of the combined normalised differential cross section to the {\sc ResBos} prediction is shown as a function of $\phist$ 
for the three $|y_Z|$ ranges in Fig.~\ref{fig:sigma_ybins}.
The measurement is also compared to predictions obtained using different MC event generators.
The PDF set  {\sc CT10} is employed in all calculations, except for {\sc Alpgen} where the CTEQ6L1 PDF set is used.
The parton-shower parameters of each MC generator are set to their default values, except for {\sc Pythia6} where a specific ATLAS re-tuning was used~\cite{ATL-PHYS-PUB-2011-009}.
The generators {\sc Alpgen}, interfaced to {\sc Herwig}, and {\sc Sherpa} provide a good description of the spectrum for $\phist >0.1$.
In particular, {\sc Sherpa} describes the data better than {\sc ResBos} over all $|y_Z|$ bins for $\phist >0.1$.
However, for $\phist < 0.1$ the deviations of {\sc Sherpa} or {\sc Alpgen} from the data are $\sim$ $5\%$, somewhat larger than those of {\sc ResBos}.
The {\sc Powheg} generator interfaced to {\sc Pythia8} is also able to describe the data to within $5\%$ over the whole $\phist$ range.

\par The effect of changing the PS tunings and algorithms interfaced to {\sc Powheg} was investigated by using {\sc Pythia6} and {\sc Herwig} interfaced to the same {\sc Powheg} NLO calculation.
These two variations give a worse description of data than {\sc Pythia8}, and deviations from data of $\sim$ $10\%$ are observed.
The \mcatnlo~generator interfaced to {\sc Herwig} does not properly describe the data for $\phist > 0.1$, and deviations from data of the order of 4--7$\%$ are observed for 
$\phist < 0.1$ depending on the $|y_Z|$ bin. 
The level of agreement between MC generators and data is very similar for comparisons at the dressed level.
%

\section{Conclusion}
\label{sec:Conclusion}

A measurement of the \phist\ distribution of $\Zg$ boson candidates in $\sqrt{s}=7$~TeV $pp$ collisions at the LHC is presented.
The data were collected with the ATLAS detector and correspond to an integrated luminosity of $4.6$~\ifb.
Normalised differential cross sections as a function of \phist\ have been measured in bins of the $Z$ boson rapidity $y_Z$ up to $\phist \sim 3$ for electron and muon pairs with an invariant mass $66\GeV<m_{\ell\ell}<116\GeV$. 
The high number of $\Zg$ boson candidates recorded permits the use of finer bins as compared to a similar study performed at the Tevatron.
The typical uncertainty achieved by the combination of electron and muon data integrated over the whole $Z$ rapidity range is below $0.5\%$ for $\phist < 0.5$ increasing to $0.8\%$ at larger $\phist$ values.

\par The cross-section measurements have been compared to resummed QCD predictions combined with fixed-order perturbative QCD calculations.
Calculations using {\sc{}ResBos} provide the best descriptions of the data.
However, they are unable to reproduce the detailed shape of the measured cross section to better than $4\%$.

\par The cross-section measurements have also been compared to predictions from different Monte Carlo 
generators interfaced to a parton shower algorithm. 
The best descriptions of the measured \phist\ spectrum are provided by {\sc{}Sherpa} and {\sc{}Powheg+Pythia8} Monte Carlo event generators.
For $\phist$ values above $0.1$, predictions from {\sc{}Sherpa} are able to reproduce the data to within $\sim 2\%$. 
The low $\phist$ part of the spectrum is, however, described less accurately than by {\sc ResBos}.
Double differential measurements as a function of $\phist$ and $y_Z$ provide valuable information for the tuning of MC generators.
None of the tested predictions is able to reproduce the detailed shape of the measured cross section within the experimental precision reached, 
which is typically lower by one order of magnitude than present theoretical uncertainties.

\global\emergencystretch = .2\hsize

\vspace{-2mm}
\section*{Acknowledgements}

%

We thank CERN for the very successful operation of the LHC, as well as the
support staff from our institutions without whom ATLAS could not be
operated efficiently.

We acknowledge the support of ANPCyT, Argentina; YerPhI, Armenia; ARC,
Australia; BMWF and FWF, Austria; ANAS, Azerbaijan; SSTC, Belarus; CNPq and FAPESP,
Brazil; NSERC, NRC and CFI, Canada; CERN; CONICYT, Chile; CAS, MOST and NSFC,
China; COLCIENCIAS, Colombia; MSMT CR, MPO CR and VSC CR, Czech Republic;
DNRF, DNSRC and Lundbeck Foundation, Denmark; EPLANET, ERC and NSRF, European Union;
IN2P3-CNRS, CEA-DSM/IRFU, France; GNSF, Georgia; BMBF, DFG, HGF, MPG and AvH
Foundation, Germany; GSRT and NSRF, Greece; ISF, MINERVA, GIF, DIP and Benoziyo Center,
Israel; INFN, Italy; MEXT and JSPS, Japan; CNRST, Morocco; FOM and NWO,
Netherlands; BRF and RCN, Norway; MNiSW, Poland; GRICES and FCT, Portugal; MERYS
(MECTS), Romania; MES of Russia and ROSATOM, Russian Federation; JINR; MSTD,
Serbia; MSSR, Slovakia; ARRS and MVZT, Slovenia; DST/NRF, South Africa;
MICINN, Spain; SRC and Wallenberg Foundation, Sweden; SER, SNSF and Cantons of
Bern and Geneva, Switzerland; NSC, Taiwan; TAEK, Turkey; STFC, the Royal
Society and Leverhulme Trust, United Kingdom; DOE and NSF, United States of
America.

The crucial computing support from all WLCG partners is acknowledged
gratefully, in particular from CERN and the ATLAS Tier-1 facilities at
TRIUMF (Canada), NDGF (Denmark, Norway, Sweden), CC-IN2P3 (France),
KIT/GridKA (Germany), INFN-CNAF (Italy), NL-T1 (Netherlands), PIC (Spain),
ASGC (Taiwan), RAL (UK) and BNL (USA) and in the Tier-2 facilities
worldwide.

\bibliographystyle{atlasBibStyle.bst}

\vspace{-2mm}
\bibliography{PhiStar_plb}


\onecolumn
\clearpage
\begin{flushleft}
{\Large The ATLAS Collaboration}

\bigskip

G.~Aad$^{\rm 48}$,
T.~Abajyan$^{\rm 21}$,
B.~Abbott$^{\rm 111}$,
J.~Abdallah$^{\rm 12}$,
S.~Abdel~Khalek$^{\rm 115}$,
A.A.~Abdelalim$^{\rm 49}$,
O.~Abdinov$^{\rm 11}$,
R.~Aben$^{\rm 105}$,
B.~Abi$^{\rm 112}$,
M.~Abolins$^{\rm 88}$
% ...
\bigskip
\\
% ...
$^{11}$ Institute of Physics, Azerbaijan Academy of Sciences, Baku, Azerbaijan\\
$^{12}$ Institut de F{\'\i}sica d'Altes Energies and Departament de F{\'\i}sica de la Universitat Aut{\`o}noma de Barcelona and ICREA, Barcelona, Spain\\
% ...
$^{21}$ Physikalisches Institut, University of Bonn, Bonn, Germany\\
% ...
$^{48}$ Fakult{\"a}t f{\"u}r Mathematik und Physik, Albert-Ludwigs-Universit{\"a}t, Freiburg, Germany\\
$^{49}$ Section de Physique, Universit{\'e} de Gen{\`e}ve, Geneva, Switzerland\\
% ...
$^{88}$ Department of Physics and Astronomy, Michigan State University, East Lansing MI, United States of America\\
% ...
$^{105}$ Nikhef National Institute for Subatomic Physics and University of Amsterdam, Amsterdam, Netherlands\\
% ...
$^{111}$ Homer L. Dodge Department of Physics and Astronomy, University of Oklahoma, Norman OK, United States of America\\
$^{112}$ Department of Physics, Oklahoma State University, Stillwater OK, United States of America\\
% ...
$^{115}$ LAL, Universit{\'e} Paris-Sud and CNRS/IN2P3, Orsay, France\\
% ...
\end{flushleft}


\end{document}